\documentclass[aps,pre,preprint,onecolumn,superscriptaddress,showpacs,floatfix]{revtex4-1}

\usepackage{graphicx}
\usepackage{amsmath}
\usepackage{amssymb}
\usepackage{epsf}
\usepackage{color}

\begin{document}

\title{Information processing in a simple one-step cascade}

\author{Mintu Nandi}
\email{nandiimintu@gmail.com}
\affiliation{Department of Chemistry, University of Calcutta, 92 A P C Road, Kolkata 700009, India}

\author{Ayan Biswas}
\email{ayanbiswas@jcbose.ac.in}
\affiliation{Department of Chemistry, Bose Institute, 93/1 A P C Road, Kolkata 700009, India}

\author{Suman K Banik}
\email{skbanik@jcbose.ac.in}
\affiliation{Department of Chemistry, Bose Institute, 93/1 A P C Road, Kolkata 700009, India}

\author{Pinaki Chaudhury}
\email{pcchem@caluniv.ac.in}
\affiliation{Department of Chemistry, University of Calcutta, 92 A P C Road, Kolkata 700009, India}

\date{\today}

\begin{abstract}
Using the formalism of information theory, we analyze the mechanism of information transduction in a simple one-step signaling cascade S$\rightarrow$X representing the gene regulatory network. Approximating the signaling channel to be Gaussian, we describe the dynamics using Langevin equations. Upon discretization, we calculate the associated second moments for linear and nonlinear regulation of the output by the input, which follows the birth-death process. While mutual information between the input and the output characterizes the channel capacity, the Fano factor of the output gives a clear idea of how internal and external fluctuations assemble at the output level. To quantify the contribution of the present state of the input to predict the future output, transfer entropy is computed. We find that higher amount of transfer entropy is accompanied by the greater magnitude of external fluctuations (quantified by the Fano factor of the output) propagation from the input to the output. We notice that low input population characterized by the number of signaling molecules S, which fluctuates in a relatively slower fashion compared to its downstream (target) species X, is maximally able to predict (as quantified by transfer entropy) the future state of the output. Our computations also reveal that with increased linear nature of the input-output interaction, all three metrics of mutual information, Fano factor and, transfer entropy achieve relatively larger magnitudes.
\end{abstract}


\maketitle


\section{Introduction}

Originally proposed by Shannon \cite{Shannon1948, Shannon1963}, information theory provides a thorough insight into the quantification, storage and communication of information \cite{Faes2015a, Faes2015b, Faes2017}. This theory has been used in diverse disciplines, one of them being signal processing motifs \cite{ Alon2006, Alon2007}, where signals are processed from an input source to an output target. Information theoretic investigations are capable of proposing a probable mechanism of signal transduction through such network motifs \cite{Mehta2009,Waltermann2011,Tareen2018}. According to this mechanism, a cell can sense the fluctuating signals originated due to either intra-cellular changes or extra-cellular changes or due to both and responded accordingly \cite{Alok2015}. Biochemical reactions are intrinsically stochastic in nature \cite{Kampen2007}, and hence cells show a strong diversity in nature due to such stochastic behavior of the biochemical networks that operate inside a single cell. Because of this stochasticity, various forms of noise are originated in these biochemical networks and consequently signal transductions are being hampered leading to a loss of information transmitted through those networks. Information theory provides the metrics to check the reliability of an organism's regulatory circuit to transmit information as well as its evolutionary fitness \cite{Donaldson2010,Rivoire2011,Taylor2007}.

Before proceeding to a discussion of our analysis, let us first introduce the notion of information transfer characterized by the quantitative term mutual information \cite{Shannon1948, Shannon1963} (MI) between the network components. MI is a probabilistic measure of the amount of information transmitted, and is estimated by quantifying the degree of dependencies between the network variables in a signaling network from a joint probability density function of those variables. For two random variables S and X forming a one-step cascade S$\rightarrow$X, MI is defined as:
\begin{equation}
\label{eq1}
I(s;x) = \sum_{s,x} p(s,x) \log_2 \frac{p(s,x)}{p(s)p(x)}
\end{equation}

\noindent where $s$ and $x$ are random variables representing biochemical species S and X respectively. $p(s,x)$ is the joint probability function of $s$ and $x$ and $p(s)$ and $p(x)$ are the marginal probability functions of $s$ and $x$ respectively. Since the base used in the logarithm is 2, the unit of the measured mutual information will be in bits. Moreover, in an alternative sense, MI is the common entropy shared by the network variables in an entropy space and signifies the average reduced uncertainty of one network variable due to the knowledge of other network variables \cite{Shannon1948, Shannon1963, Cover1991}. MI, a non-negative quantity, is a symmetric measure of correlation between stochastic variables of the system and its value becomes zero if the network variables are independent of each other \cite{Cover1991}.

Though MI provides a quantified overlap of the information content between the network variables, it does not contain any dynamical as well as directional information \cite{Schreiber2000}. While analyzing a network motif, it becomes a central task to find out the direction as well as quantification of information transmission. Thomas Schreiber \cite{Schreiber2000} proposed a quantitative measurement of information flow, named transfer entropy (TE) \cite{Bossomaier2016, Prokopenko2014} containing the desired properties which are not within the scope of MI measurement. Unlike MI, measurement of TE takes transition probabilities into account. For the same two random variables, the transition probability $p(x_{i+h}|x_i,s_i)$ has to be considered where $x_{i+h}$ is the value of $x$ which is $h$ step in future from $x_i$. For a discrete-time process, the prediction horizon $h$ is considered to be 1 \cite{Wibral2013} as we have used in the present work (see Section~II). TE, the amount of information transmitted in a particular direction under the given scheme of network motif, signifies how reliably one can have the information about the future state of the target output from the present state of the input source provided the present state of the output is known. Therefore, it is a history-based measurement. A novel application of TE was demonstrated by Bauer \textit{et.al.} \cite{Bauer2007}. They showed that TE detects the path of propagation of disturbance among the process variables in a continuous chemical scheme. For this purpose, they took two industrial case studies and identified the root sources of the disturbance correctly. They found significant values of TE associated with those sources of disturbance and these values of TE satisfied a significance level which exceeds a threshold value of TE ascertained for the system. A recent study has also revealed important connections between information theory and thermodynamics by exploring the effect of time reversal on transfer entropy \cite{Spinney2016}.

Keeping the ideas of MI and TE and their broad utility to understand varied chemical networks and their operation, we, in our present effort, have studied the one-step cascade (OSC) motif with a view to understand how Fano factor, MI and TE can be used to analyze information propagation in the network which faces stochastic environment. Here, it is imperative to understand the real biological phenomenon which has been mathematically modelled by an OSC motif. The biochemical species represented here by S and X may well be envisaged as two gene products, the former is responsible for the production of the later \cite{Alon2007}. Gene product S, in our case, acts as a transcription factor (TF), more accurately, an activator binding to the promoter region of the gene of X and thereby, up-regulating production of X. Single-cell experiments categorically suggest these binding events, apart from the production and degradation events, to be inherently noisy \cite{Kaern2005,Paulsson2005}. These activator molecules originate and diffuse through the cytoplasm and after arriving in the nucleus, get bound to the promoter region of the target gene and initiate transcribing and translating the information content of the source gene. The TFs often get unbound from the promoter site and with time again get attached, thereby adding noise into the system. These microscopic details are encapsulated in the coarse-grained systemic parameters, whose effect on the statistical metrics (i.e., Fano factor, MI and, TE), this present study wants to capture. The motifs found in the transcription control networks of \textit{E. coli} and \textit{S. cerevisiae} can be depicted as composed of further basic interaction units of OSC. Hence, information theoretic profiling of OSC seems to be of utmost importance. This type of analysis also opens avenues of comparing different metrics on equal footing to find out some common features, if present.

Our treatment of this motif has been done both as an analytic treatment as well as numerical investigation. In the next section, we describe in detail the methodology used and then go on to present the results with accompanying discussions. We end our presentation by making the concluding remarks as to how the information theory especially transfer entropy has helped in the understanding of OSC motif.


\section{The Model}

\begin{figure}[!t]
\includegraphics[scale=0.7]{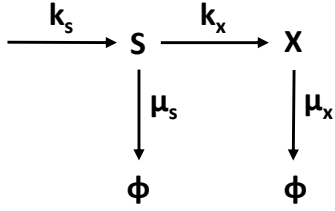}
\caption{The OSC motif. $k_s$ and $k_x$ are the synthesis rate of $S$ 
and $X$, respectively. The degradation constants of the same components 
are given by $\mu_s$ and $\mu_x$. The rate constants $\mu_s$, $k_x$ 
and $\mu_x$ are expressed in sec$^{-1}$ while $k_s$ is expressed in 
(molecules/$V$) sec$^{-1}$, $V$ being the unit effective cellular volume.
While drawing the OSC motif we considered only the linear 
forms of the functions $f_s(s) = k_s$ and $f_x(s,x) = k_x s$ [see 
Eqs.~(\ref{eq2}-\ref{eq3}) and discussion before Eq.~(\ref{eq12})].}
\label{f1}
\end{figure}

We consider signal transmission through an OSC motif. The OSC 
motif consists of an input S and an output X (see Fig.~\ref{f1}).
The time evolution of S and X can be written using a set of coupled Langevin 
equations
\begin{eqnarray}
\label{eq2}
\frac{ds}{dt} & = & f_s(s) - \mu_s s + \xi_s(t), \\
\label{eq3}
\frac{dx}{dt} & = & f_x(s,x)-\mu_x x + \xi_x(t),
\end{eqnarray}


\noindent
where $s$ and $x$ are the copy numbers of S and X, respectively. Here, copy number stands for the number of molecules of biochemical species and is expressed in molecules/$V$ where $V$ is the unit effective cellular volume. In Eqs.~(\ref{eq2}-\ref{eq3}), $f_s (s)$ and $f_x (s,x)$ represent production terms associated with S and X, respectively. Both the synthesis terms $f_s (s)$ and $f_x (s,x)$ could be, in general, nonlinear in nature. In Fig.~\ref{f1}, however, we show the linear case only. Explicit forms of synthesis terms are given when we consider specific cases.
The noise processes $\xi_s (t)$ and $\xi_x (t)$ are Gaussian distributed with statistical properties 
\cite{Gillespie2000,Elf2003,Paulsson2004,Tanase2006,Kampen2007}
\begin{eqnarray*}
\langle \xi_{s} (t) \rangle & = & \langle \xi_{x} (t) \rangle = 0, \\
\nonumber
\langle \xi_{s} (t)\xi_{s}(t^\prime) \rangle & = & 
[ f_s  (\langle s \rangle) + \mu_{s} \langle s \rangle ]
{\delta^\prime} (t-t^\prime), \\
\nonumber
\langle \xi_{x} (t) \xi_{x} (t^\prime) \rangle & = & 
[ f_x (\langle s \rangle, \langle x \rangle) + \mu_x \langle x \rangle ]
{\delta^\prime} (t-t^\prime), \\
\langle \xi_{s}(t) \xi_{x} (t^\prime) \rangle & = & 
\langle \xi_{x} (t) \xi_{s} (t^\prime) \rangle = 0 .
\end{eqnarray*}

\noindent where ${\delta^\prime}$ is the Dirac delta function. In the above expressions 
of noise correlations, $\langle \cdots \rangle$ stands for ensemble average.
We note here that, at steady state
$f_s (\langle s \rangle) + \mu_{s} \langle s \rangle = 2 \mu_{s} \langle s \rangle$
and
$f_x (\langle s \rangle, \langle x \rangle) + \mu_{x} \langle x \rangle = 2 \mu_{x} \langle x \rangle$
as synthesis and degradation terms balance each other
\cite{Elf2003,Paulsson2004,Tanase2006,Kampen2007}.

\subsection{Analytical calculations}

To solve the Langevin equations (\ref{eq2}-\ref{eq3}), we perform discretization
\begin{eqnarray}
\label{eqn1}
s_t & = & f_s (s_{t-1}) \Delta t + (1-\beta_1) s_{t-1} + \epsilon^s_t, \\
\label{eqn2}
x_t & = & f_x (s_{t-1},x_{t-1}) \Delta t + (1-\beta_2) x_{t-1} + \epsilon^x_t,
\end{eqnarray}

\noindent with
$\beta_1 = \mu_s \Delta t$, $\beta_2 = \mu_x \Delta t$,
$\epsilon^s_t = \xi_s(t) \sqrt{\Delta t}$ and
$\epsilon^x_t = \xi_x(t) \sqrt{\Delta t}$. Here,
$\Delta t$ is the small time grid. We note that, the discrete set of Eqs.~(\ref{eqn1}-\ref{eqn2}) is an approximated form of continuous-time process, Eqs.~(\ref{eq2}-\ref{eq3}) under the assumption that $\Delta t \rightarrow$ 0. The approximation we make here has been discussed in detail by Barnett et al. \cite{Barnett2009} in connection to the discretized approximation of the continuous-time process. We further note that in our numerical calculation we also adopt the discretized version of the Langevin equations (see Section~IIB). The set of discretized equations are the starting point of our subsequent analysis.
After performing ensemble average at steady state, Eqs.~(\ref{eqn1}-\ref{eqn2}) become
\begin{eqnarray*}
\label{eqn3}
\langle s_t \rangle\ & = & f_s ( \langle s_{t-1} \rangle ) \Delta t
+ (1-\beta_1) \langle s_{t-1} \rangle, \\
\label{eqn4}
\langle x_t \rangle\ & = & f_x ( \langle s_{t-1} \rangle, \langle x_{t-1} \rangle ) \Delta t
+ (1-\beta_2) \langle x_{t-1} \rangle.
\end{eqnarray*}

\noindent The fluctuations associated with $s_t$ and $x_t$ around mean value $\langle s \rangle$ and $\langle x \rangle$, respectively, are given by
\begin{eqnarray}
\label{eqn5}
\delta s_t = s_t - \langle s_t \rangle 
& = & [ f_s (s_{t-1}) - f_s ( \langle s_{t-1} \rangle ) ] \Delta t \nonumber \\
&& + (1 - \beta_1) \delta s_{t-1} + \epsilon^s_t, \\
\label{eqn6}
\delta x_t = x_t - \langle x_t \rangle 
& = & [ f_x (s_{t-1},x_{t-1}) - f_x ( \langle s_{t-1} \rangle, \langle x_{t-1} \rangle ) ] 
\Delta t \nonumber \\
&& + (1 - \beta_2) \delta x_{t-1} + \epsilon^x_t.
\end{eqnarray}

\noindent
We now define 
\begin{eqnarray*}
f^{\prime}_s ( \langle s_{t-1} \rangle ) & = &
\lim_{\delta s_{t-1} \rightarrow 0}
\frac{f_s (s_{t-1}) - f_s ( \langle s_{t-1} \rangle )}{\delta s_{t-1}}, \\
f^{\prime}_{x,s} ( \langle s_{t-1} \rangle, \langle x_{t-1} \rangle ) & = &
\lim_{\delta s_{t-1} \rightarrow 0}
\frac{f_x (s_{t-1}, x_{t-1}) - f_x ( \langle s_{t-1} \rangle, \langle x_{t-1} \rangle )}{\delta s_{t-1}},
\end{eqnarray*}

\noindent
and rewrite Eqs.~(\ref{eqn5}-\ref{eqn6}) as
\begin{eqnarray}
\label{eqn7}
\delta s_t & = & [ f^{\prime}_s ( \langle s_{t-1} \rangle ) \Delta t 
+ (1 - \beta_1) ] \delta s_{t-1} + \epsilon^s_t, \\
\label{eqn8}
\delta x_t & = & f^{\prime}_{x,s} ( \langle s_{t-1} \rangle, \langle x_{t-1} \rangle )
\Delta t \delta s_{t-1}
+ (1 - \beta_2) \delta x_{t-1} + \epsilon^x_t.
\end{eqnarray}

\noindent Eqs.~(\ref{eqn7}-\ref{eqn8}) can be rewritten in the matrix form\cite{Barnett2009,Lizier2012}
\begin{equation}
\label{eq4}
\delta \mathbf{C_t} = \mathbf{B} \delta \mathbf{C_{t-1}} + \mathbf{N}, 
\end{equation}

\noindent
with
\begin{eqnarray*}
\delta \mathbf{C_t} = 
\left ( \begin{array}{c}
\delta s_t \\
\delta x_t
\end{array} \right ), 
\delta \mathbf{C_{t-1}} = 
\left ( \begin{array}{c}
\delta s_{t-1} \\
\delta x_{t-1}
\end{array} \right ), \\
\mathbf{B} = 
\left ( \begin{array}{cc}
f^{\prime}_s ( \langle s \rangle ) \Delta t + 1 - \beta_1 & 0 \\
f'_{x,s}(\langle s\rangle,\langle x\rangle) \Delta t & 1 - \beta_2
\end{array} \right ) \; {\rm and,} \;
\mathbf{N} = 
\left ( \begin{array}{c}
\epsilon_t^s \\
\epsilon_t^x
\end{array} \right ).
\end{eqnarray*}

\noindent 
While writing Eq.~(\ref{eq4}) we have used $\langle s_{t-1} \rangle = \langle s_t \rangle \equiv \langle s \rangle$ and $\langle x_{t-1} \rangle = \langle x_t \rangle \equiv \langle x \rangle$ at steady state. 
Now, multiplying $\delta \mathbf{C_t}$ with its transpose $\delta \mathbf{C_t^T}$ ($\mathbf{T}$ stands for transpose) followed by an averaging (ensemble), we arrive at
\begin{equation}
\label{eqn9}
\langle \delta \mathbf{C_t} \delta \mathbf{C_t^T} \rangle
= \mathbf{B} \langle \delta \mathbf{C_{t-1}} \delta \mathbf{C_{t-1}^T} \rangle
\mathbf{B^T} + \langle \mathbf{N N^T} \rangle.
\end{equation}

\noindent
While writing Eq.~(\ref{eqn9}) we have set  $\mathbf{B} \langle \delta \mathbf{C_{t-1}} \mathbf{N^T} \rangle$ and $\langle \mathbf{N} \delta \mathbf{C_{t-1}^T} \rangle \mathbf{B^T}$ equals to zero as $\delta \mathbf{C_{t-1}}$ and $\mathbf{N}$ are uncorrelated to each other. When written explicitly Eq.~(\ref{eqn9}) becomes
\begin{eqnarray}
\label{eqn10}
\left (
\begin{array}{cc}
\langle \delta s_t \delta s_t \rangle & \langle \delta s_t \delta x_t \rangle \\ 
\langle \delta x_t \delta s_t \rangle & \langle \delta x_t \delta x_t \rangle
\end{array}
\right )
& = &
\left ( \begin{array}{cc}
f^{\prime}_s ( \langle s \rangle ) \Delta t + 1 - \beta_1 & 0 \\
f'_{x,s}(\langle s\rangle,\langle x\rangle) \Delta t & 1 - \beta_2
\end{array} \right )
\left (
\begin{array}{cc}
\langle \delta s_{t-1} \delta s_{t-1} \rangle & \langle \delta s_{t-1} \delta x_{t-1} \rangle \\ 
\langle \delta x_{t-1} \delta s_{t-1} \rangle & \langle \delta x_{t-1} \delta x_{t-1} \rangle
\end{array}
\right )
\nonumber \\
&& \times
\left ( \begin{array}{cc}
f^{\prime}_s ( \langle s \rangle ) \Delta t + 1 - \beta_1 & 0 \\
f'_{x,s}(\langle s\rangle,\langle x\rangle) \Delta t & 1 - \beta_2
\end{array} \right )^T
+
\left (
\begin{array}{cc}
\langle \epsilon^s_t \epsilon^s_t \rangle & \langle \epsilon^s_t \epsilon^x_t \rangle \\ 
\langle \epsilon^x_t \epsilon^s_t \rangle & \langle \epsilon^x_t \epsilon^x_t \rangle
\end{array}
\right )
\end{eqnarray}

\noindent At steady state, it is approximated that the mean square deviations are time-invariant and hence statistically identical at all time points. Also, keeping in view the symmetry properties of the covariances, we have
\begin{eqnarray*}
\langle \delta s_t \delta s_t \rangle 
& = & \langle \delta s_{t-1} \delta s_{t-1} \rangle 
= \langle \delta s^2_t \rangle = \sigma_s^2, \\
\langle \delta x_t \delta x_t \rangle 
& = & \langle \delta x_{t-1} \delta x_{t-1} \rangle 
= \langle \delta x^2_t \rangle = \sigma_x^2, \\
\langle \delta s_t \delta x_t \rangle 
& = & \langle \delta x_t \delta s_t \rangle 
= \langle \delta s_{t-1} \delta x_{t-1} \rangle \\
& = & \langle \delta x_{t-1} \delta s_{t-1} \rangle = \sigma_{sx}^2.
\end{eqnarray*}

\noindent 
The respective noise strengths at steady state are
\begin{eqnarray*}
\langle \epsilon^s_t \epsilon^s_t \rangle 
& = & \langle \xi_s (t) \xi_s (t) \rangle \Delta t 
= 2 \mu_s \langle s \rangle \Delta t, \\
\langle \epsilon^x_t \epsilon^x_t \rangle 
& = & \langle \xi_x (t) \xi_x (t) \rangle \Delta t 
= 2 \mu_x \langle x \rangle \Delta t, \\
\langle \epsilon^s_t \epsilon^x_t \rangle 
& = & \langle \xi_s (t) \xi_x (t) \rangle \Delta t = 0, \\
\langle \epsilon^x_t \epsilon^s_t \rangle 
& = & \langle \xi_x (t) \xi_s (t) \rangle \Delta t = 0.
\end{eqnarray*}

\noindent 
Hence Eq.~(\ref{eqn10}) becomes
\begin{eqnarray}
\label{eqn11}
\left (
\begin{array}{cc}
\sigma_s^2 & \sigma_{sx}^2 \\ 
\sigma_{sx}^2 & \sigma_x^2
\end{array}
\right )
& = &
\left ( \begin{array}{cc}
f^{\prime}_s ( \langle s \rangle ) \Delta t + 1 - \beta_1 & 0 \\
f'_{x,s}(\langle s\rangle,\langle x\rangle) \Delta t & 1 - \beta_2
\end{array} \right )
\left (
\begin{array}{cc}
\sigma_s^2 & \sigma_{sx}^2 \\ 
\sigma_{sx}^2 & \sigma_x^2
\end{array}
\right )
\left ( \begin{array}{cc}
f^{\prime}_s ( \langle s \rangle ) \Delta t + 1 - \beta_1 & 0 \\
f'_{x,s}(\langle s\rangle,\langle x\rangle) \Delta t & 1 - \beta_2
\end{array} \right )^T
\nonumber \\
&& +
\left (
\begin{array}{cc}
2 \mu_s \langle s \rangle \Delta t & 0 \\ 
0 & 2 \mu_x \langle x \rangle \Delta t
\end{array}
\right )
\end{eqnarray}

\noindent
Eq.~(\ref{eqn11}) can be written as 
\begin{equation}
\label{eq5}
\mathbf{A = B A B^T + M},
\end{equation}

\noindent
where
\begin{eqnarray*}
\mathbf{A} =
\left (
\begin{array}{cc}
\sigma_s^2 & \sigma_{sx}^2 \\ 
\sigma_{sx}^2 & \sigma_x^2 
\end{array}
\right ),
\mathbf{M} =  
\left (
\begin{array}{cc}
2 \mu_s \langle s \rangle \Delta t & 0 \\ 
0 & 2 \mu_x \langle x \rangle \Delta t
\end{array}
\right ).
\end{eqnarray*}

\noindent 
The solution of Eq.~(\ref{eq5}) provides the expressions of variance and covariance associated with the components S and X.
The corresponding expressions for the second moments at time point $t$ are
\begin{eqnarray}
\label{eq6}
\sigma_s^2 & = & \frac{
2 \beta_1 \langle s \rangle
}{
1 - [f^{\prime}_s ( \langle s \rangle ) \Delta t + 1 - \beta_1]^2
}, \\
\label{eq7}
\sigma_{sx}^2 & = & \frac{
\sigma_s^2 
[f'_{x,s}(\langle s\rangle,\langle x\rangle)  \Delta t] 
[f^{\prime}_s ( \langle s \rangle ) \Delta t + 1 - \beta_1]
}{
1 - [f^{\prime}_s ( \langle s \rangle ) \Delta t + 1 - \beta_1] (1 - \beta_2)
}, \\
\label{eq8}
\sigma_x^2 & = & 
\frac{
2 \beta_2 \langle x \rangle 
+  [f'_{x,s}(\langle s\rangle,\langle x\rangle) \Delta t]^2 \sigma_s^2 
+ 2 [ f'_{x,s}(\langle s\rangle,\langle x\rangle) \Delta t] (1-\beta_2) \sigma_{sx}^2
}{
\beta_2 (2 - \beta_2)
}.
\end{eqnarray}

\noindent Now, replacing $t$ by $(t+1)$ in Eq.~(\ref{eq4}) and multiplying both side of the resulting equation by $\delta \mathbf{C_t}^T$ and subsequently taking average we get \cite{Barnett2009,Lizier2012}
\begin{equation}
\label{eq9}
\mathbf{A_{t+1,t} = BA},
\end{equation}

\noindent where
\begin{eqnarray*}
\mathbf{A} & = & \langle \delta \mathbf{C_t} \delta \mathbf{C_t^T} \rangle, \nonumber \\
\mathbf{A_{t+1,t}} & = & \langle \delta \mathbf{C_{t+1}} \delta \mathbf{C_t^T} \rangle. \nonumber
\end{eqnarray*}

\noindent Using the explicit form of $\delta \mathbf{C_t}$ one may also write
\begin{equation}
\nonumber
\mathbf{A_{t+1,t}}=\left ( \begin{array}{cc}
\langle \delta s_{t+1}\delta s_t\rangle & \langle \delta s_{t+1}\delta x_t\rangle \\ 
\langle \delta x_{t+1}\delta s_t\rangle & \langle \delta x_{t+1}\delta x_t\rangle
\end{array} \right ).
\end{equation}

\noindent At this point, we note that Eq.~(\ref{eq9}) represents one-lag covariance for a discrete-time process. However, the same equation is a result of discrete-time approximation of the continuous-time process given by Eqs.~(\ref{eq2}-\ref{eq3}). For a continuous-time process, the accurate representation of the one-lag covariance involves a matrix exponential \cite{Barnett2009}. In the present work, however, we adopt the discrete-time approximation of the continuous-time process to compute both theoretical and numerical results.
Now, comparing both sides of Eq.~(\ref{eq9}), the expressions for one-lag covariances associated with S and X are obtained
\begin{eqnarray}
\label{eq10}
\sigma^2_{x_{t+1},s_t} & = & \langle \delta x_{t+1}\delta s_t\rangle \nonumber \\
& =  & f'_{x,s}(\langle s\rangle,\langle x\rangle) \Delta t \sigma^2_s 
+ (1-\beta_2)\sigma^2_{sx}, \\
\label{eq11}
\sigma^2_{x_{t+1},x_t} & = & \langle \delta x_{t+1}\delta x_t\rangle \nonumber \\
& =  & f'_{x,s}(\langle s\rangle,\langle x\rangle) \Delta t \sigma^2_{sx} 
+ (1-\beta_2)\sigma^2_x.
\end{eqnarray}

\noindent We now write the explicit forms of the functions $f_s(s)$ and $f_x(s,x)$ keeping in mind the nature of the biochemical interaction.
For linear interaction, we have \cite{Alok2015}
\begin{eqnarray*}
\nonumber
f_s(s) = k_s, \; f_x(s,x) = k_x s.
\end{eqnarray*}

\noindent
and for nonlinear case \cite{Bintu2005,Bialek2008,Tkacik2008b,Tkacik2008c,Ziv2007}
\begin{eqnarray*}
f_s(s) = k_s, \; f_x(s,x) = k_x\left(\frac{s^n}{K^n+s^n}\right)
\end{eqnarray*}

\noindent
where $K$ is the activation coefficient having the same dimension of $s$. $K$ takes care of threshold concentration of S required to activate the expression of X. $n$ takes care of co-operative interaction among different S, commonly known as Hill coefficient \cite{Alon2007}. In both linear and nonlinear cases, we assume simplest form of $f_s (s) = k_s$ that dictates Eq.~(\ref{eq2}) to follow Poisson process. It is important to note that, however, $f_s (s)$ may be a nonlinear function of S dictated by autoregulation.
On the other hand, depending on the nature of interaction we assumed two different forms of $f_x (s,x)$. For $K \gg s$, one may write $f_x (s,x) \approx (k_x/K^n) s^n$ which is same as in the linear case (for $n=1$), but with a scaled value of $k_x$. We further note that the unit of $k_x$ is different in both cases. In linear case unit of $k_x$ is sec$^{-1}$ while in nonlinear case it becomes (molecules/$V$) sec$^{-1}$.
In the rest of our analysis, we have used $n=1$ and $K = \langle s \rangle$. $n=1$ takes care of binding of a single S in the promoter of X. Multiple binding and resultant cooperativity are taken care of by $n > 1$, which we have not incorporated in the present work. Here, $K = \langle s \rangle$ signifies half-maximal expression of X due to S \cite{Alon2007}.

In the Gaussian framework, the MI between S and X can be expressed as
\begin{equation}
\label{eq12}
I(s;x)=\frac{1}{2} \log_2
\left ( \frac{\sigma_s^2}{\sigma_{s|x}^2}
\right ),
\end{equation}

\noindent where the conditional variance $\sigma_{s|x}^2 = \sigma_s^2 - (\sigma_{sx}^4 / \sigma_x^2)$ \cite{Barrett2015}. In the above expression, the unit of MI is bits.
TE is interpreted as a quantitative measure of the direction of information flow. In our context, TE measures the information flow from the input S to the output X. TE can be defined as the information contributed from the present state of S (at time $t$) to the future state of X (at time $t$+1), given the knowledge of the present state of X (at time $t$). In principle, this implies that $x_{t+1}$ can be predicted with greater accuracy when the knowledge of both $s_t$ and $x_t$ is available as opposed to the knowledge of $x_t$ alone. So, it measures the extent of influence of the present state of S on the future state of X. Hence we express TE using MI as follows \cite{Barrett2015}:
\begin{equation}
\label{eq13}
\tau_{s\rightarrow x} = I(x_{t+1};s_t|x_t) =  I(x_{t+1};s_t,x_t) - I(x_{t+1};x_t).
\end{equation}

\noindent
Here, the first term signifies the mutual information between the present states of S, X and the future state of X, whereas the second term implies the mutual information between the present and future states of X. In terms of second moments obtained so far the analytical expression of transfer entropy becomes (see Appendix A)
\begin{equation}
\label{eq14}
\tau_{s\rightarrow x} = \frac{1}{2} \log_2 \left (
\frac{\det \Delta_1 \det \Delta_2}{\sigma^2_x \det \Delta_3}
\right ),
\end{equation}

\noindent with
\begin{eqnarray*}
\Delta_1 & = &
\left (
\begin{array}{cc}
\sigma^2_s & \sigma^2_{sx} \\ 
\sigma^2_{sx} & \sigma^2_x
\end{array}
\right ),
\Delta_2 = 
\left (
\begin{array}{cc}
\sigma^2_x & \sigma^2_{x_{t+1},x_t} \\ 
\sigma^2_{x_{t+1},x_t} & \sigma^2_x
\end{array}
\right ), \\
\Delta_3 & = & 
\left (
\begin{array}{ccc}
\sigma^2_x & \sigma^2_{x_{t+1},x_t} & \sigma^2_{x_{t+1},s_t} \\
\sigma^2_{x_{t+1},x_t} & \sigma^2_x & \sigma^2_{sx} \\
\sigma^2_{x_{t+1},s_t} & \sigma^2_{sx} & \sigma^2_s
\end{array}
\right ).
\end{eqnarray*}

\subsection{Numerical calculations}

\begin{table}
\caption{List of kinetic parameter $\gamma$, duration of Gillespie simulation $t_f$ and discrete-time step $\Delta t$. Note that $\gamma = \mu_s/\mu_x$ is the ratio of degradation rate of S and X, respectively. The corresponding values of $\mu_s$ and $\mu_x$ are given within parenthesis for each parametric value of $\gamma$.
}
\begin{ruledtabular}
\begin{tabular}{ccc}
$\gamma~(= \mu_s/\mu_x)$ & $t_f$ (sec) & $\Delta t$ (sec) \\
\hline
0.1~(= 1/10) & 100 & $3.6 \times 10^{-3}$ \\
1~(= 10/10) & 20 & $7.5 \times 10^{-4}$ \\
10~(= 10/1) & 50 & $1 \times 10^{-3}$\\
\end{tabular}
\end{ruledtabular}
\end{table}

We numerically simulate the nonlinear network using Gillespie algorithm \cite{Gillespie1976, Gillespie1977} and quantify relevant metrics for OSC at steady state. We show that simulated data agree well with the analytical results. In the numerical simulation, we have generated time series of the input and output network components. In these time series, when the populations of the components reach limiting values and do not suffer any considerable change with time, the components are taken to attain the steady state. To make sure of this, the simulation is carried out for a significant length of time (say $t_f$), which is mentioned for each of the parametric scenarios in Table~I. We have generated $10^6$ independent trajectories. From each trajectory, we have discarded time samples up to $t = t_f - 2$ and collected the last two time samples i.e., at $t = t_f - 1$ and $t = t_f$, where $t_f$ is the final time up to which the simulation is carried out. We note that the time difference between $t_f - 1$ and $t_f$ differs in each trajectory as the algorithm uses a random waiting time between successive reactions. Thus accurate measurement of the duration of transient kinetics is difficult to calculate from Gillespie time series. At this point it is important to mention that the discrete-time size can affect the calculation of transfer entropy \cite{Spinney2017,Barnett2017}. In Table~I, we thus provide the values of the discrete-time step $\Delta t$ used for theoretical calculation corresponding to each parametric value of $\gamma$ where $\gamma = \mu_s/\mu_x$. The collected data points are used to evaluate various marginal and joint probability density functions (PDFs). These PDFs, in turn, are used for calculating the associated statistical properties, e.g. Fano factor,  mutual information and transfer entropy associated with the motif.

Following Schreiber \cite{Schreiber2000}, the expression of TE in terms of PDF's is as follows
\begin{equation}
\label{eq15}
\tau_{s \rightarrow x} = \sum p(x_{t+h},x_t^{(k)},s_t^{(l)})\log_2 \frac{p(x_{t+h}|x_t^{(k)},s_t^{(l)})}{p(x_{t+h}|x_t^{(k)})}.
\end{equation}

\noindent In this paper, we calculate transfer entropy using this discrete-time formula and co-plot with the corresponding analytical results obtained from a discrete-time process (Eqs.~(\ref{eqn1}-\ref{eqn2})). Besides, the prediction horizon $h$ is taken to be unity and the embedding length, $k=l=1$ \cite{Wibral2013} which, is computationally easy to handle.
The transition probabilities can be reduced to joint PDF's as follows \cite{Gardiner2009}
\begin{eqnarray*}
p(x_{t+1}|x_t,s_t) & = &\frac{p(x_{t+1},x_t,s_t)}{p(x_t,s_t)}, \\
p(x_{t+1}|x_t) & = & \frac{p(x_{t+1},x_t)}{p(x_t)}.
\end{eqnarray*}

\noindent 
Therefore,
\begin{equation}
\label{eq16}
\tau_{s \rightarrow x} = \sum p(x_{t+1},x_t,s_t)\log_2 \frac{p(x_{t+1},x_t,s_t)p(x_t)}{p(x_{t+1},x_t)p(x_t,s_t)}.
\end{equation}

\noindent The marginal and joint PDF's are evaluated using the Kernel method \cite{Epanechnikov1969, Silverman1986}. In this method, for a given sample of $n$ observations $(y_1,y_1,...,y_n)$ the definition of the empirical PDF at any point $y$ is given by
\begin{equation}
\label{eq17}
p(y) = \frac{1}{n} \sum_{i=1}^n \frac{1}{h_y} K_e\left(\frac{y-y_i}{h_y}\right), 
\end{equation}

\noindent where $K_e$ is the Kernel function used in this estimation and $h_y$ is the optimal Kernel width fitted for the Kernel used. A Kernel function is a symmetric function which satisfies the following conditions
\begin{eqnarray*}
\int K_e(a) da = 0, \int aK_e(a) da = 0 \; {\rm and} \int a^2K_e(a) da \neq 0.
\end{eqnarray*}

\noindent We have used Epanechnikov Kernel \cite{Epanechnikov1969} as an optimum Kernel function which has the maximum efficiency of 1 over the other kernels given in the literature \cite{Epanechnikov1969, Silverman1986}. The Epanechnikov kernel is defined as
\begin{eqnarray*}
 K_e(a)  =  \left\{
\begin{array}{ccc}
  \frac{1}{3\sqrt{5}}\left(1-\frac{a^2}{5}\right) & {\rm for} & |a| < \sqrt{5}, \\
   0  & {\rm otherwise} 
\end{array}
\right .
\end{eqnarray*}

\noindent This estimation for univariate case is extended to the multivariate random variable where the empirical PDF for the $k$-variate random variable $Z_i=Z(z_1^{(i)},z_2^{(i)},...,z_k^{(i)})$ where ${i=1,2,...,n}$, is defined as \cite{Silverman1986}
\begin{equation}
\label{eq18}
p(z_1,z_2...,z_k) = \frac{1}{n} \sum_{i=1}^n \prod_{j=1}^k 
\frac{1}{h_{z_j}}K_{e_j} \left ( \frac{z_j-z_j^{(i)}}{h_{z_j}} \right ).
\end{equation}

\noindent The optimal Kernel widths for Epanechnikov Kernel for the case of each individual univariate data $z_j$ is given by Silverman \cite{Silverman1986} are $h_{z_j} = \kappa \sigma_{z_j} n^{-1/5}$ with $\kappa = (40\sqrt{\pi})^{1/5}$ and $j=1,2,...,k$. Here $\sigma_{z_j}$ is the standard deviation of the $n$ realizations of the univariate random variable $z_j$.


\section{Results and Discussion}

In this section, we present a comparative study of the OSC motif as shown in Fig.~\ref{f1}. We investigate the information transmission through the OSC motif under the effect of various parameters associated with the motif. TE has been taken as an efficient metric to analyse the directed information transmission through the cascade.
The OSC motif (Fig.~\ref{f1}) is a signaling cascade in which the signal S regulates the gene to form protein X. To this end we consider here the linear and nonlinear form of the functions $f_s(s)$ and $f_x(s,x)$, respectively, i.e., $f_s(s) = k_s$ and $f_x(s,x) = k_x ( s/(K+s) )$. Now for this kinetic scheme, one can numerically compute the protein distribution, $p(x)$ at steady state. Here, we aim to observe the effect of variation of signal strength (i.e., $\langle s\rangle$) on the steady state protein distribution, $p(x)$.

\begin{figure}[!t]
\includegraphics[width=1.0\columnwidth,angle=0]{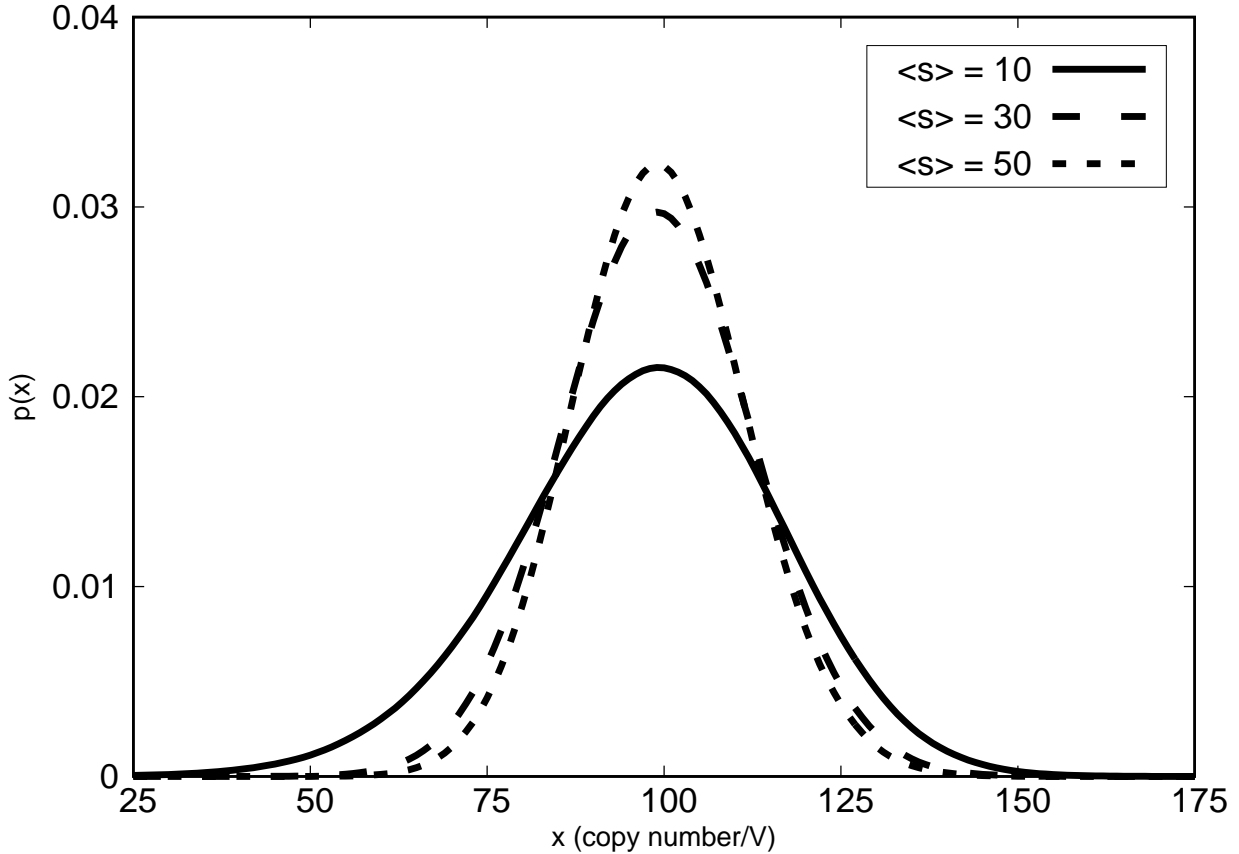}
\caption{Probability distribution of protein at steady-state. The mean population of the output $\left\langle x\right\rangle=100$ molecules/V and the value of $\gamma=\mu_s/\mu_x = 0.1$. The other relevant parameters associated with the motif are governed by the relations: $k_s=\mu_s \langle s\rangle$ and $k_x=\mu_x \langle x\rangle \left[(K +\langle s\rangle)/\langle s\rangle \right]$, with $K=\langle s\rangle$. The profiles are generated from numerical simulation using Gillespie's algorithm \cite{Gillespie1976,Gillespie1977}.}
\label{f2}
\end{figure}

Fig.~\ref{f2} shows the steady-state protein distribution for three different steady-state populations of signal e.g.,  $\left\langle s\right\rangle=10,30$ and $50$. This figure reveals that for smaller population of input signal (e.g., $\left\langle s\right\rangle=10$) the distribution is broader and with increasing $\left\langle s\right\rangle$, it becomes sharply peaked. The variance is a measure of the spread of the protein distribution around its mean value. So, in principle, with a continued increment of $\left\langle s\right\rangle$, the variance associated with the distribution will be significantly reduced. At this state, the mean population of the signal will be appreciably high, and one can expect the system to show a much-reduced level of fluctuations in the associated protein population.

\begin{figure}[!t]
\includegraphics[width=1.0\columnwidth,angle=0]{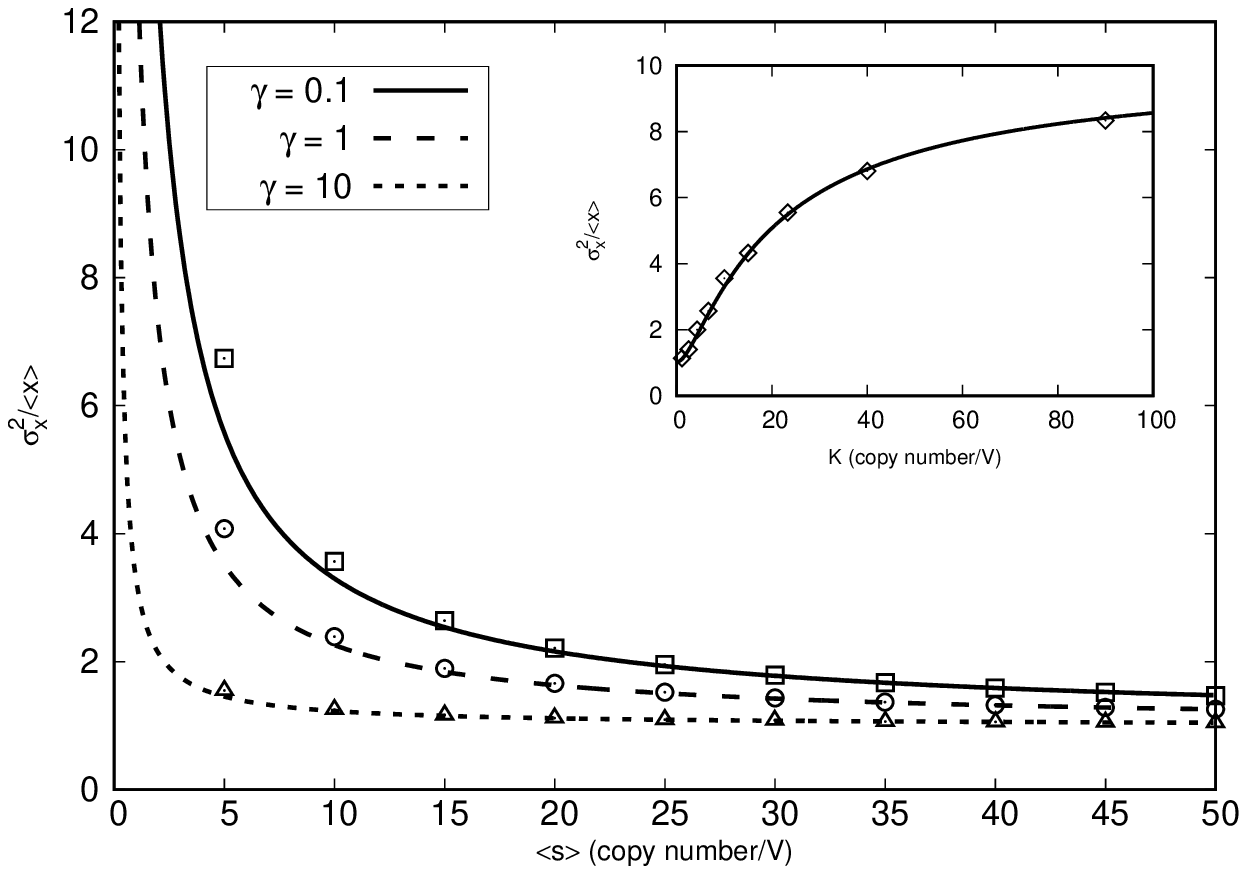}
\caption{Variation of Fano factor ($\sigma_x^2/\langle x\rangle$) as a function of mean population of the signal $\langle s\rangle$ expressed in molecules/V. The mean output population is kept fixed at $\langle x\rangle=100$ molecules/V. We choose $\Delta t =3.6 \times10^{-3},~7.5 \times10^{-4} ~\rm{and}, 10^{-3} sec$ for $\gamma =0.1,1 ~\rm{and}, 10$ respectively. The other relevant parameters associated with the motif are governed by the relations: $k_s=\mu_s \langle s\rangle$ and $k_x=\mu_x \langle x\rangle \left[(K + \langle s\rangle)/\langle s\rangle\right]$, where $K=\langle s\rangle$. $\bf{Inset}$: Profile of Fano factor as a function of $K$ for $\gamma=0.1$. The mean population of the signal and the output are $\langle s\rangle=10$ and $\langle x\rangle=100$ respectively, both are expressed in molecules/V. Here, $\Delta t = 4.2 \times 10^{-3} sec$. Other relevant parameters are set according to the relations: $k_s=\mu_s \langle s\rangle$ and $k_x=\mu_x \langle x\rangle \left[(K+\langle s\rangle)/\langle s\rangle\right]$. In both the profiles, the lines are drawn from theoretical calculations and the symbols are generated from numerical simulation using Gillespie's algorithm \cite{Gillespie1976,Gillespie1977}.}
\label{f3}
\end{figure}

Fig.~\ref{f3} shows an exponentially decreasing nature of the Fano factor as a function of $\left\langle s\right\rangle$, where Fano factor is a measure of relative fluctuations associated with the corresponding component. To calculate Fano factor, we neglect $\Delta t^2$ in Eqs.~(\ref{eq6}-\ref{eq8}) as we use $\Delta t = 10^{-3}-10^{-4}$ sec. As a result of it, we obtain the expression of Fano factor associated with the protein level (see Appendix B)
\begin{equation}
\label{eq19}
F(x) = 1+ 
\frac{\langle x\rangle}{\langle s\rangle (1+\gamma) [1+\langle s\rangle /K]^2},
\end{equation}

\noindent where $\gamma = \mu_s/\mu_x$.

Fig.~\ref{f3} represents a gradual decrease in relative fluctuations in protein level with the increase in the mean population of the signal. At the high mean population of the same, Fano factor associated with protein distribution gets reduced significantly, implying an accumulation of very low level of fluctuations in protein pool. With increasing $\gamma$ value, the Fano factor is also diminished which is due to the higher relative fluctuations of input compared to the output. The rate of fluctuations is recorded by the degradation rates of the system components. Hence, the separation of time scales makes the output unable to sense the input. The inset in Fig.~\ref{f3}, depicts how the profile of Fano factor changes as one modifies the interaction from nonlinear to linear form. Fixing $\langle s\rangle$=10 and $\langle x\rangle$=100, as we increase $K$ the interaction between S and X becomes more linear, and a hyperbolic increase in Fano factor can be observed. To explain this trend, we resort to Eq.~(\ref{eq19}), which clearly shows that the second term in Fano factor expression increases with increasing $K$ value.
The variance of the protein distribution measures the magnitude of fluctuations associated with protein itself but cannot explain how fluctuations influence the information transfer through the cascade. To understand the underlined mechanism, we take resort to the information-theoretic formalism as proposed by Shannon \cite{Shannon1948, Shannon1963}.

\begin{figure}[!t]
\includegraphics[width=1.0\columnwidth,angle=0]{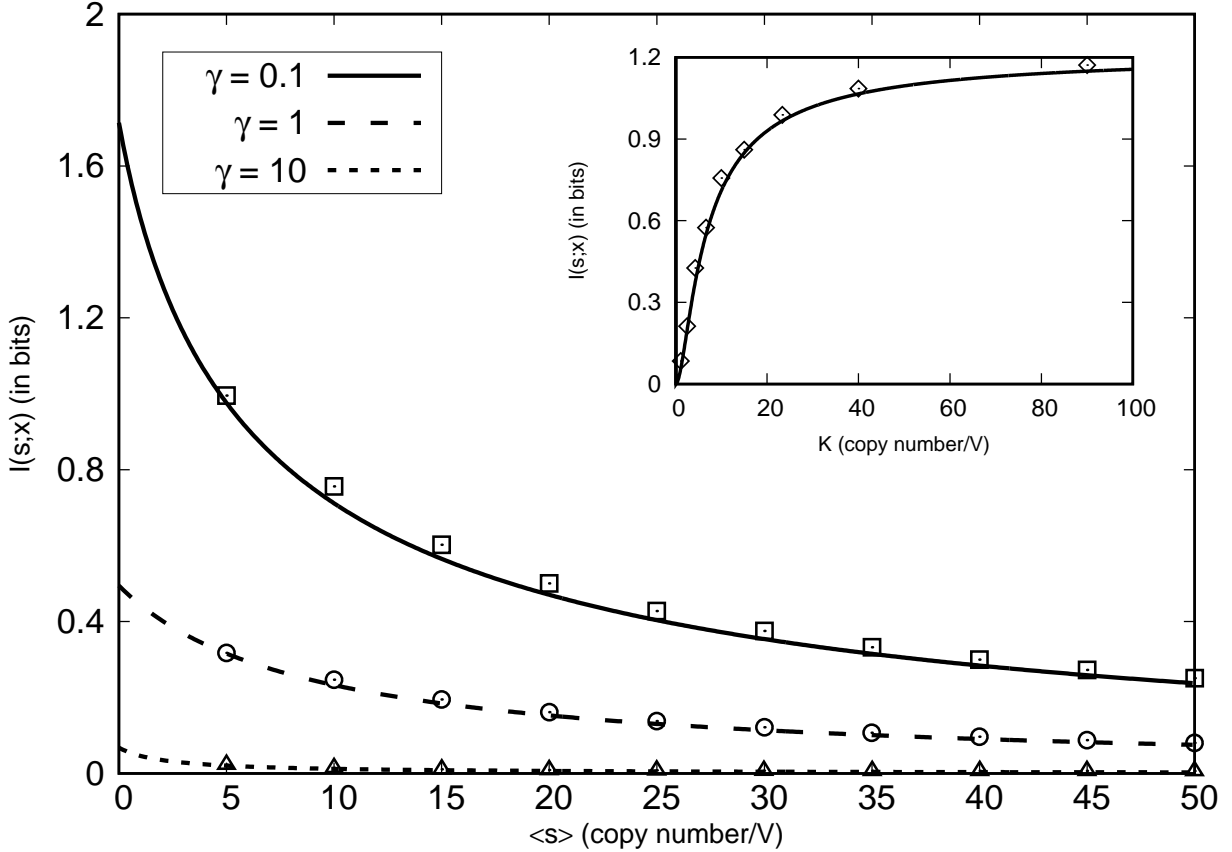}
\caption{Variation of mutual information ($I(s;x)$) as a function of mean population of the signal $\langle s\rangle$ expressed in molecules/V. The mean output population is kept fixed at $\langle x\rangle=100$ molecules/V. We choose $\Delta t =3.6 \times10^{-3},~7.5 \times10^{-4} ~\rm{and}, 10^{-3} sec$ for $\gamma =0.1,1 ~\rm{and}, 10$ respectively. The other relevant parameters associated with the motif are governed by the relations: $k_s=\mu_s \langle s\rangle$ and $k_x=\mu_x \langle x\rangle \left[(K + \langle s\rangle)/\langle s\rangle \right]$, where $K=\langle s\rangle$. $\bf{Inset}$: Profile of mutual information as a function of $K$ for $\gamma=0.1$. The mean population of the signal and the output are $\langle s\rangle=10$ and $\langle x\rangle=100$ respectively, both are expressed in molecules/V. Here, $\Delta t = 4.2 \times 10^{-3} sec$. Other relevant parameters are set according to the relations: $k_s=\mu_s \langle s\rangle$ and $k_x=\mu_x \langle x\rangle \left[(K + \langle s\rangle)/\langle s\rangle \right]$. In both the profiles, the lines are drawn from theoretical calculations and the symbols are generated from numerical simulation using Gillespie's algorithm \cite{Gillespie1976,Gillespie1977}.}
\label{f4}
\end{figure}

According to Shannon's definition of MI, it is an average measure of reduced uncertainty for a random variable when the knowledge of the other random variable is available \cite{Shannon1948, Shannon1963}. In our context, MI between the two species, the signaling component S and the protein X, measures the reduced uncertainty in protein level by knowing S and vice versa. In Fig.~\ref{f4}, it has been observed that MI decreases gradually with increasing $\left\langle s\right\rangle$. A low value of MI signifies that the two components become less correlated and hence the information about the input signal which is sensed by the response, gets diminished. The reduction in MI with increasing $\gamma$ can be explained by separation of time scales in the OSC motif. Inset of Fig.~\ref{f4} also conveys the fact that an increasing amount of linearity in the system increases MI. To account for such behavior, one looks back at Eqs.~(\ref{eq7}-\ref{eq8}) which have higher values for linear interaction compared to its nonlinear situation. This can be realized by noting that $f'_{x,s}(\langle s\rangle,\langle x\rangle)_{linear} > f'_{x,s}(\langle s\rangle,\langle x\rangle)_{nonlinear}$ as 
$f'_{x,s}(\langle s\rangle,\langle x\rangle)_{linear} = k_x$ and
$f'_{x,s}(\langle s\rangle,\langle x\rangle)_{nonlinear} = k_x K/(K + \langle s \rangle)^2$.

Along with MI which is a symmetric correlation measure between S and X, we also analyse the information transfer along the cascade in terms of TE which will help us to harness the amount of information propagation in the cascade. MI detects only mutually overlapped fluctuations spaces between S and X. If we know the fluctuations space of S, we can predict the same for X and vice versa. But TE measures the fluctuations space of X by the knowledge of S, but the reverse is not possible since X does not stimulate S in OSC. As a result, unlike MI, TE makes a better account of the fluctuations level in X. It is also imperative to investigate any existent similarity between the nature of variations of MI, Fano factor and, TE.

\begin{figure}[!t]
\includegraphics[width=1.0\columnwidth,angle=0]{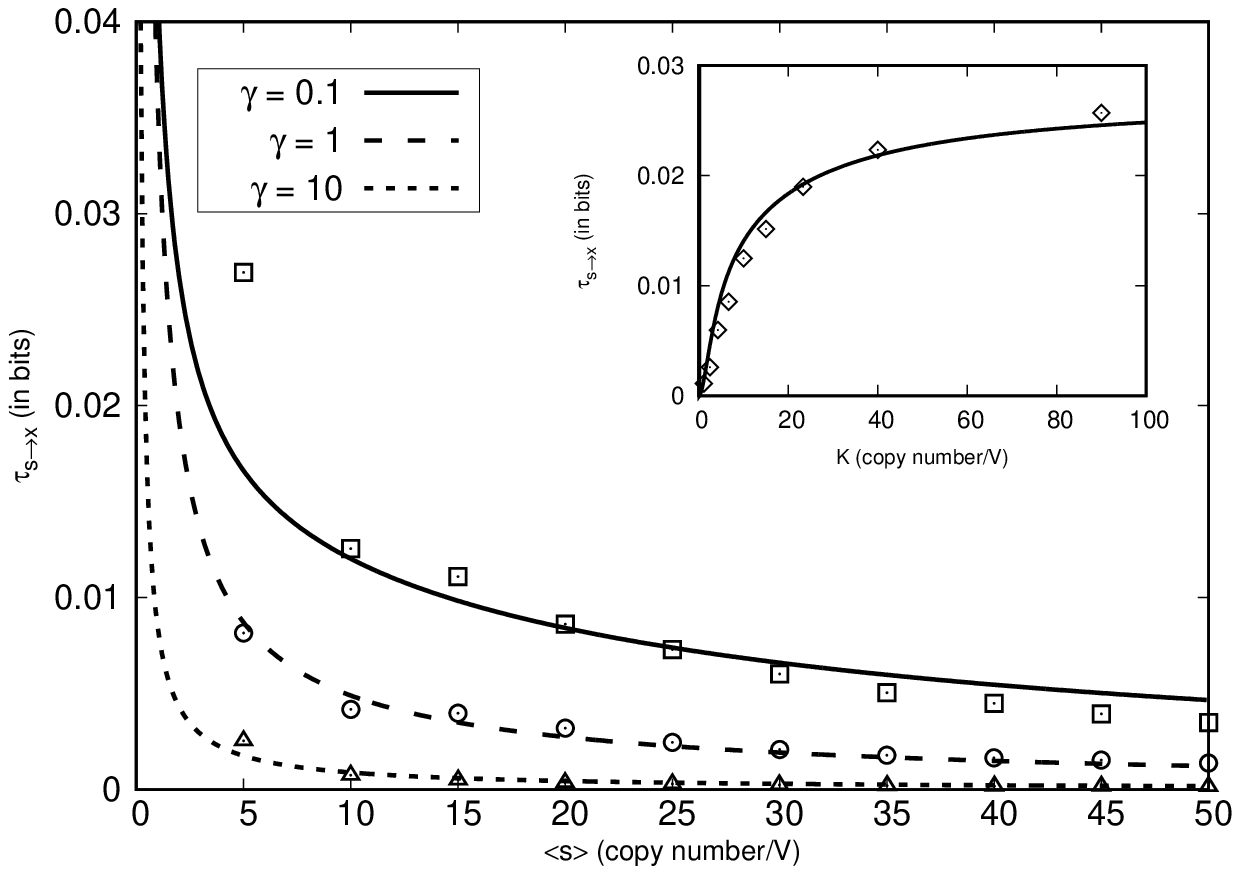}
\caption{Variation of transfer entropy ($\tau_{s\rightarrow x}$) as a function of mean population of the signal $\langle s\rangle$ expressed in molecules/V. The mean output population is kept fixed at $\langle x\rangle=100$ molecules/V. We choose $\Delta t =3.6 \times10^{-3},~7.5 \times10^{-4} ~\rm{and}, 10^{-3} sec$ for $\gamma =0.1,1 ~\rm{and}, 10$ respectively. The other relevant parameters associated with the motif are governed by the relations: $k_s=\mu_s \langle s\rangle$ and $k_x=\mu_x \langle x\rangle \left[(K+\langle s\rangle)/\langle s\rangle\right]$, where $K=\langle s\rangle$. $\bf{Inset}$: Profile of transfer entropy as a function of $K$ for $\gamma=0.1$. The mean population of the signal and the output are $\langle s\rangle=10$ and $\langle x\rangle=100$ respectively, both are expressed in molecules/V. Here, $\Delta t = 4.2 \times 10^{-3} sec$. Other relevant parameters are set according to the relations: $k_s=\mu_s \langle s\rangle$ and $k_x=\mu_x \langle x\rangle \left[(K+\langle s\rangle)/\langle s\rangle\right]$. In both the profiles, the lines are drawn from theoretical calculations and the symbols are generated from numerical simulation using Gillespie's algorithm \cite{Gillespie1976,Gillespie1977}.}
\label{f5}
\end{figure}

Here, we analyse the variation of TE with the steady-state population of the signal observed under different parametric situations, namely three different values of $\gamma$. Fig.~\ref{f5} shows the variation of TE with  $\left\langle s\right\rangle$ for $\gamma =10.0,1.0$ and $0.1$. There is a sharp exponential decay for $\gamma=10.0$, but the sharpness of the decay profile decreases as we go to lower values of $\gamma$. This decrease indicates a greater amount of information transmitted through the network. At signaling level, the relative fluctuations around the signaling component is determined by the Fano factor i.e., $F(s)=\sigma_s^2/\left\langle s\right\rangle$. From Eq.~(\ref{eq6}), by neglecting $\Delta t ^2$ terms one gets, $\sigma_s^2 =  \left\langle s\right\rangle $ (see Appendix B) and hence $F(s)=1$. So, the Fano factor of S remains always the same at unity, and this is because S exhibits a linear birth and death process and hence it follows Poisson statistics. Since the relative fluctuations around the signaling component are always the same independent of the variation in $\gamma$ value, one question may arise here that, what should be the noise source in S which influences the information transfer along the cascade? Since $F(s)$ remains constant, the signaling component suffers fluctuations only due to its relaxation time scale. This time scale act as an extrinsic noise source at the protein level in the sense that the protein level serves as the system of interest \cite{Elowitz2002}. This protein population also has fluctuations associated with itself. In fact, noise (both intrinsic and extrinsic) substantially reduces the information transduced from the signaling source. So, when $\gamma=10.0$, $\mu_s \gg \mu_x$ which indicates faster fluctuations of the signal in comparison with that of protein population, leading to a greater extrinsic noise generated at the signaling level. Hence one can say that protein species cannot sense the rapid population fluctuations of the signal \cite{Alok2015}. So, it can be expected to have a lower magnitude of TE indicating a lower amount of information transduced from signal to the protein which forms the response. On the other hand, when $\gamma=1.0$, i.e., $\mu_s = \mu_x$, the relaxation time scales of the signal and the response species are matched. Consequently, in contrast to the former case, the protein can effectively sense the population changes of the signal by having a lesser extrinsic noise produced at the signaling level. Therefore, a greater extent of information is expected to be transduced from signal to the protein, leading to a greater magnitude of TE. For the case of $\gamma=0.1$, the relaxation time of the signal is much higher than that of the protein, and hence the response can effectively sense the much slower fluctuations of the signal with ease. Consequently, protein level can accumulate more information about the signal fluctuations in this relaxation time scale limit in comparison with the other two cases. As a result, a much greater amount of information is processed from the signal to the protein. These three different variations of TE for three different magnitudes of $\gamma$ are shown in Fig.~\ref{f5}.

For any particular value of $\gamma$, TE decreases with increasing $\left\langle s\right\rangle$. This increase in $\left\langle s\right\rangle$ is obtained by increasing $k_s$ while keeping $\left\langle x\right\rangle$ fixed at 100. So, this indicates that a large population of the signaling component allows transduction of little amount of information to downstream protein population. As we go to the low population level of the signaling molecule, its information transduction capacity increases giving a higher value of TE in comparison to the former case. To account for this, we take note of Eq.~(\ref{eq19}) giving the expression for the Fano factor of the protein population. On the right-hand side of this equation, the first term (unit value) arises due to the internal fluctuations (INL) caused by the variations in concentration of protein whereas the second term is due to the external fluctuations (EFL) originated at the signaling level
\begin{equation}
\label{eq20}
{\rm EFL} =\frac{\langle x\rangle}{\langle s\rangle (1+\gamma) [1+\langle s\rangle /K]^2}.
\end{equation}

\noindent
According to Bauer $et~al$ \cite{Bauer2007}, disturbance produced at a certain process variable propagates from that variable to the other variables in a chemical process. This disturbance contains information about the relative changes happening at that process variable, and this gets propagated along the path in the direction of the process flow. So, in this system of interest, information transmission in a certain direction is nothing but the propagation of fluctuations which contain information about the changes of the signal strength, along with that direction. TE does its work to identify the direction of fluctuations propagation and finds the extent of information transduced along that direction. Greater the propagation of fluctuations, greater will be the transduction of information which in turn produces a greater value of TE. The EFL in the protein level is incorporated due to the propagation of fluctuations from the signaling component to protein. So, it is understood that lesser propagation of fluctuations is reflected in the low value of EFL. With increasing $\left\langle s\right\rangle$, the value of EFL decreases indicating a decrease in the extent of fluctuations propagating along the cascade, and consequently, we get a diminished magnitude of TE. Fig.~\ref{f5} clearly depicts this diminishing behavior of TE with $\left\langle s\right\rangle$. Again, for a particular value of $\left\langle s\right\rangle$, if we change $\gamma$ from $10$ to $0.1$ as described in the previous paragraph, the magnitude of EFL increases and hence we get an increase in TE because of the propagation of higher degree of fluctuations along the cascade. Fig.~\ref{f5} includes the inset panel which shows an increasing pattern of TE with increasing $K$. Taking a hint from the relationship of $I(s;x)$ with $K$, it is evident that the time-lagged MI terms that constitute TE also increase with increasing $K$. This can be indirectly observed by noting the increasing trend of $\det \Delta_1$, $\det \Delta_2$, $\det \Delta_3$ and $\sigma_x^2$ with increasing $K$.


\section{Conclusion}

Our work concentrates on the propagation of fluctuations in a one-step cascade using the tools of Shannon information theory. Within the stochastic framework, we have used metrics of mutual information, Fano factor and transfer entropy to analyze the complex correlation pattern of two random variables having a unidirectional regulation between them. These random variables which adhere to Gaussian approximation can well represent different biochemical species in a complex network. The key findings of our study are

\begin{itemize}

\item Fluctuations in the population level of the output species can be separated into two different categories namely internal and external fluctuations. 

\item Low copy number of input variable contributes a higher degree of fluctuations into the system. The external fluctuations originate at the level of input due to the low population can be characterized by the increasing width of the output probability distribution as we decrease the input population. 

\item The external fluctuations are also modulated by the separation of the relaxation time scale, which exits between the input and the output. Its signature is prominent in the profiles of Fano factor of the output, mutual information and transfer entropy. 

\item The profiles for all of these metrics follow the hyperbolic trend with increasing input population. Additionally, when the ratio of input to output relaxation rates ($\gamma$) is increased, all these profiles show reduced magnitudes. So, it signifies with fast variation in the input population level relative to that of the output population, and the output species fails to sense the input fluctuations in a reliable way. 

\item We noticed that with the low population of input and a smaller ratio of relaxation time scales between the input and output, the input species could have better predictive power over the output species and this predictive power sharply increases, thereby establishing a strong causal connection towards the future state of the response.

\item It is also revealed that a more significant amount of fluctuations are propagated from upstream to downstream species under the scenario mentioned in the previous point.

\item We obtained results that demonstrate a linear system produces higher values for Fano factor, mutual information and transfer entropy compared to a nonlinear system.

\end{itemize}

To summarize, low concentration and slower relaxation rate of TF relative to that of the target (X) allow the cascade to propagate significant amount of fluctuations in the protein level and thereby enhance the predictability about the fluctuations space associated with the protein level. According to Fraser and co-workers \cite{Fraser2004}, in the synthesis of essential and complex forming proteins, low level of fluctuations are associated with the output. Hence, if the concentration of TF is high and rate of relaxation of the same is faster relative to that of the gene product (i.e., protein), these essential proteins can accumulate the low level of fluctuations. Again, in some circumstances fluctuations in the protein concentration are advantageous. In that case, the upstream regulatory gene product (TF) having low concentration fluctuates relatively slower with respect to protein to produce a large amount of fluctuations in the protein pool.

We believe, our study on the fluctuations propagation in OSC motif using the tool of transfer entropy paves the way to analyze more complicated networks given the advantage of the modular property of those networks which appear in various biological phenomena. The results of our work may also contribute as a starting point towards designing synthetic biological circuits.

 
\begin{acknowledgments}
\noindent
Mintu Nandi and Ayan Biswas are thankful to UGC (22/06/2014(i)EU-V) and 
Bose Institute, Kolkata, respectively for research fellowship. Financial support from Council of Scientific and Industrial Research (CSIR), India
[01(2771)/14/EMR-II] is thankfully acknowledged.\end{acknowledgments}


\appendix


\section{The mutual information terms in Eq.(\ref{eq13})}

The mutual information terms in Eq.~(\ref{eq13}) can be written in terms of 
Shannon entropy as
\begin{eqnarray*}
I(x_{t+1};s_t,x_t) & = & H(x_{t+1}) + H(s_t,x_t) - H(x_{t+1},s_t,x_t), \\
I(x_{t+1},x_t) & = & H(x_{t+1}) + H(x_t) - H(x_{t+1},x_t).
\end{eqnarray*}

\noindent 
The Shannon entropy terms in the above equations can be written as
\begin{eqnarray*}
\nonumber H(s_t,x_t) = H(s_t)+ H(x_t|s_t),
\end{eqnarray*}

\noindent where,
\begin{eqnarray*}
H(s_t) & = & \frac{1}{2}\log_2 (\sigma_{s}^2) + \frac{1}{2}\log_2(2\pi e), \\
H(x_t|s_t) & = & \frac{1}{2}\log_2 ( \sigma_{x_t|s_t}^2 ) + \frac{1}{2}\log_2(2\pi e), \\
\sigma_{x_t|s_t}^2 & = & \sigma_x^2-\frac{\sigma^4_{sx}}{\sigma^2_s}.
\end{eqnarray*}

\noindent Therefore,
\begin{equation}
H(s_t,x_t) = \frac{1}{2}\log_2\left[\det \Delta_1 \right]+\log_2(2\pi e), 
\label{eqA1}
\end{equation}

\noindent where,
$\Delta_1 = 
\left (
\begin{array}{cc}
\sigma^2_s & \sigma^2_{sx} \\ 
\sigma^2_{sx} & \sigma^2_x
\end{array}
\right )$.

\begin{eqnarray*}
\nonumber H(x_{t+1},x_t) = H(x_{t+1})+H(x_t|x_{t+1}),
\end{eqnarray*}

\noindent where,
\begin{eqnarray*}
H(x_{t+1}) & = & \frac{1}{2}\log_2 (\sigma_x^2)) + \frac{1}{2}\log_2(2\pi e), \\
H(x_t|x_{t+1}) & = & \frac{1}{2}\log_2 (\sigma_{x_t|x_{t+1}}^2) + \frac{1}{2}\log_2(2\pi e), \\
\sigma_{x_t|x_{t+1}}^2 & = & \sigma^2_x - \frac{\sigma^4_{x_{t+1},x_t}}{\sigma^2_x}.
\end{eqnarray*}

\noindent Therefore,
\begin{equation}
H(x_{t+1},x_t) = \frac{1}{2}\log_2\left[\det \Delta_2 \right]+\log_2(2\pi e),
\label{eqA2}
\end{equation}

\noindent where, 
$\Delta_2 = 
\left (
\begin{array}{cc}
\sigma^2_x & \sigma^2_{x_{t+1},x_t} \\ 
\sigma^2_{x_{t+1},x_t} & \sigma^2_x
\end{array}
\right )$.

\begin{equation}
\nonumber H(x_{t+1},s_t,x_t) = H(x_{t+1}) + H(s_t|x_{t+1}) + H(x_t|x_{t+1},s_t),
\end{equation}

\noindent where,
\begin{eqnarray*}
\nonumber H(x_{t+1}) & = & \frac{1}{2}\log_2[\sigma_x^2]+\frac{1}{2}\log_2(2\pi e), \\
\nonumber H(s_t|x_{t+1}) & = & \frac{1}{2}\log_2[\sigma_{s_t|x_{t+1}}^2]+\frac{1}{2}\log_2(2\pi e), \\
\nonumber H(x_t|x_{t+1},s_t) & = & \frac{1}{2}\log_2[\sigma_{x_t|x_{t+1},s_t}^2]+\frac{1}{2}\log_2(2\pi e), \\
\nonumber \sigma_{s_t|x_{t+1}}^2 & = & \sigma^2_s - \frac{\sigma^4_{x_{t+1},s_t}}{\sigma^2_x},\\
\nonumber \sigma_{x_t|x_{t+1},s_t}^2 & = & \sigma_{x_t}^2 - 
\left (
\begin{array}{cc}
\sigma_{x_{t+1},x_t}^2 & \sigma_{s_t,x_t}^2
\end{array}
\right )
\left (
\begin{array}{cc}
\sigma_{x_{t+1}}^2 & \sigma_{s_t,x_{t+1}}^2 \\ 
\sigma_{x_{t+1},s_t}^2 & \sigma_{s_t}^2
\end{array}
\right )^{-1} \\
\nonumber && \times 
\left (
\begin{array}{c}
\sigma_{x_t,x_{t+1}}^2 \\ 
\sigma_{x_t,s_t}^2
\end{array}
\right ).
\end{eqnarray*}

\noindent After completing the matrix multiplication, the above partial variance gives
\begin{equation}
\nonumber \sigma_{x_t|x_{t+1},s_t}^2 = \sigma^2_x - \frac{\sigma^2_s\sigma^4_{x_{t+1},x_t} - 2\sigma^2_{sx}\sigma^2_{x_{t+1},s_t}\sigma^2_{x_{t+1},x_t} + \sigma^2_x\sigma^4_{sx}}{\sigma^2_s\sigma^2_x - \sigma^4_{x_{t+1},s_t}}.
\end{equation}

\noindent Therefore,
\begin{equation}
H(x_{t+1},s_t,x_t) = \frac{1}{2}\log_2\left[\det \Delta_3 \right]+\frac{3}{2}\log_2(2\pi e),
\label{eqA3}
\end{equation}

\noindent where, 
$\Delta_3 =
\left (
\begin{array}{ccc} 
\sigma^2_x & \sigma^2_{x_{t+1},x_t} & \sigma^2_{x_{t+1},s_t} \\
\sigma^2_{x_{t+1},x_t} & \sigma^2_x & \sigma^2_{sx} \\
\sigma^2_{x_{t+1},s_t} & \sigma^2_{sx} & \sigma^2_s
\end{array}
\right ).$

\begin{equation}
H(x_t) = \frac{1}{2}\log_2[\sigma^2_x]+\frac{1}{2}\log_2(2\pi e).
\label{eqA4}
\end{equation}

\noindent Hence, analytic expressions of the mutual information terms in Eq.[\ref{eq13}] are expressed as follows
\begin{eqnarray}
I(x_{t+1};s_t,x_t) & = & \frac{1}{2}\log_2\left(\frac{\sigma_x^2 \det \Delta_1}{\det \Delta_3}\right) \label{eqA5}, \\
I(x_{t+1};x_t) & = & \frac{1}{2}\log_2\left(\frac{\sigma_x^4}{\det \Delta_2}\right) \label{eqA6}.
\end{eqnarray}


\section{Calculation of Fano factor given in Eq.~(\ref{eq19})}

For nonlinear interaction (with $n=1$) we have $f^{\prime}_s (\langle s \rangle) = 0$ 
and
$f^{\prime}_{x,s} (\langle s \rangle, \langle x \rangle) = k_x K/ (K + \langle s \rangle)^2$.
Now using Eq.~(\ref{eq6}), we have
\begin{eqnarray}
\label{eqb1}
\sigma_s^2 & = & \frac{
2 \beta_1 \langle s \rangle
}{
1 - [f^{\prime}_s ( \langle s \rangle ) \Delta t + 1 - \beta_1]^2
}, \nonumber \\
& \approx & \langle s \rangle,
\end{eqnarray}

\noindent where we have neglected the terms with $\Delta t^2$, i.e., 
$(1 - \beta_1)^2 \approx 1 - 2 \beta_1$, as $\beta_1 = \mu_s \Delta t$.
Similarly, using Eq.~(\ref{eq7}), $\beta_1 = \mu_s \Delta t$ and 
$\beta_2 = \mu_x \Delta t$ we write
\begin{eqnarray}
\label{eqb2}
\sigma_{sx}^2 & = & \frac{
\sigma_s^2 
[f'_{x,s}(\langle s\rangle,\langle x\rangle)  \Delta t] 
[f^{\prime}_s ( \langle s \rangle ) \Delta t + 1 - \beta_1]
}{
1 - [f^{\prime}_s ( \langle s \rangle ) \Delta t + 1 - \beta_1] (1 - \beta_2)
}, \nonumber \\
& \approx & \sigma_s^2
\frac{k_x K \Delta t}{(K + \langle s \rangle)^2}
\times \frac{1}{(\mu_s + \mu_x) \Delta t}, \nonumber \\
& = & \frac{k_x K \langle s \rangle}{(\mu_s + \mu_x) (K + \langle s \rangle)^2}.
\end{eqnarray}

\noindent On a similar note, from Eq.~(\ref{eq8}) we have
\begin{eqnarray}
\label{eqb3}
\sigma_x^2 & = & 
\frac{
2 \beta_2 \langle x \rangle 
+  [f'_{x,s}(\langle s \rangle,\langle x \rangle) \Delta t]^2 \sigma_s^2 
+ 2 [ f'_{x,s}(\langle s \rangle,\langle x \rangle) \Delta t] (1-\beta_2) \sigma_{sx}^2
}{
\beta_2 (2 - \beta_2)
}, \nonumber \\
& \approx &
2 \mu_x \Delta t \left [ \langle x \rangle + 
\frac{
\langle x \rangle^2
}{
\langle s \rangle (1 + \gamma) 
\left ( 1 + \frac{\langle s \rangle}{K} \right )^2 
}
\right ] 
\times
\frac{1}{2 \mu_x \Delta t }, \nonumber \\
& = &
\langle x \rangle + 
\frac{
\langle x \rangle^2
}{
\langle s \rangle (1 + \gamma) 
\left ( 1 + \frac{\langle s \rangle}{K} \right )^2
},
\end{eqnarray}

\noindent
where $\gamma = \mu_s / \mu_x$. Now, using the definition of 
Fano factor we have the expression given in Eq.~(\ref{eq19})
\begin{eqnarray*}
F (x) = \frac{\sigma_x^2}{\langle x \rangle} =
1 + 
\frac{
\langle x \rangle
}{
\langle s \rangle (1 + \gamma) 
\left ( 1 + \frac{\langle s \rangle}{K} \right )^2 
}.
\end{eqnarray*}



\end{document}